\documentclass[12pt]{article}
\usepackage{amsfonts,graphics}
\newcommand{\ZZ}{{\mathbb Z}}
\newcommand{\sfrac}[2]{{\textstyle{\frac{#1}{#2}}}}

\begin{document}
\newpage
\pagestyle{empty}
\vfill
\begin{center}

{\Large \textbf{Symmetry and Codon Usage Correlations} 

\vspace{2mm}

\textbf{in the Genetic Code}}

\vspace{10mm}

{\large L. Frappat, P. Sorba}

\vspace{4mm}

{\em Laboratoire de Physique Th\'eorique LAPTH, URA 1436,}
\\
{\em Chemin de Bellevue, BP 110,\\ F-74941 Annecy-le-Vieux, France}
\\
{\em E-mail:} \texttt{frappat(sorba)@lapp.in2p3.fr} 

\vspace{7mm}

{\large A. Sciarrino}

\vspace{4mm}

{\em Dipartimento di Scienze Fisiche, Universit\`a di Napoli
       ``Federico II''}
\\
{\em and I.N.F.N., Sezione di Napoli, Italy}
\\
{\em Mostra d'Oltremare Pad. 20, 80125 Napoli, Italy}
\\
{\em E-mail:} \texttt{sciarrino@na.infn.it}
\end{center}

\vfill

\begin{abstract}
The ratios of the codon usage in the quartets and sextets for the 
vertebrate series exhibit a correlated behaviour which fits naturally in 
the framework of the crystal basis model of the genetic code \cite{FSS}.  
Moreover the observed universal behaviour of these suitably normalized 
ratios can be easily explained.
\end{abstract}

\vfill
\vfill

\rightline{LAPTH-709/98}
\rightline{physics/9812041}
\rightline{December 1998}

\newpage
\phantom{blankpage}
\newpage
\pagestyle{plain}
\setcounter{page}{1}
\baselineskip=17pt

%%%%%%%%%%%%%%%%%%%%%%%%%%%%%%%%%%%%%%%%%%%%%%%%%%%%%%%%%%%%%%%%%%%%%%%

\section{Introduction}
 
It is a well known and intriguing fact that, in the genetic code, 64 codons 
code the biosynthesis of 20 amino-acids (a.a.)  with a structure in 
multiplets reported in Table \ref{tablerep}. \\
It is also a well known and, at our knowledge, unexplained fact that the 
frequency rate of usage (codon usage) of the different codons inside a 
multiplet is not the same.  \\
It is the aim of this paper to emphasize for the vertebrate series a 
correlation in the codon usage, in the quartets and sextets, which is 
naturally explained in the framework of the mathematical model of the 
genetic code recently proposed by the authors \cite{FSS}.  Moreover we put 
in evidence an universal function behaviour connected with the codon usage, 
which also finds a justification in the model.  \\
In Sec.  2 we recall the essential features of the model and in Sec.  3 we 
present the analysis of the codon usage for a set of biological sequences 
in the vertebrate series \cite{data}.

\section{The crystal basis model}
 
\subsection{The crystal basis}

Let us briefly recall some properties of the crystal basis \cite{Kashi}.  
We limit ourselves to the case of ${\cal U}_{q}(sl(2))$, but such basis 
exists for any finite dimensional representation of ${\cal U}_{q 
\rightarrow 0}({\cal G})$ , ${\cal G}$ being any (semi)-simple classical 
Lie algebra.  The crystal basis has the nice property, for ${\cal 
U}_{q}(sl(2))$, that in the limit $q \rightarrow 0$
\begin{eqnarray} 
\widetilde{J}_{+} \, u_{k} &= & u_{k+1} \quad \mathrm{for} \,\, 0 \le k < 
2J \\
\widetilde{J}_{-} \, u_{k} &= & u_{k-1} \quad \mathrm{for} \,\, 0 < k \le 
2J \\
\widetilde{J}_{3} \, u_{k} &= & (k-J) \, u_{k} \quad \mathrm{for} \,\, 0 
\le k \le 2J
\end{eqnarray} 
and
\begin{equation} 
\widetilde{J}_{+} \, u_{2J} = \widetilde{J}_{-} \, u_{0} = 0 
\end{equation} 
where the operators $\widetilde{J}_{\pm }$ are a redefinition, using an 
element of the center, of the generators $J_{\pm}$ of ${\cal 
U}_{q}(sl(2))$, $\widetilde{J}_{3} = J_{3}$, and $u_{k}$ are the basis 
vectors of the irreducible representation labelled by $J$ ($J$ being an 
integer or half-integer).  The labels of the irreducible representation are 
connected to the eigenvalues of the ``Casimir'' operator $C$:
\begin{equation} 
C = (\widetilde{J}_{3})^{2} + \frac{1}{2} \sum_{n \in {\ZZ_+}} \sum_{k=0}^n 
(\widetilde{J}_{-})^{n-k} (\widetilde{J}_{+})^n (\widetilde{J}_{-})^k \,.
\end{equation} 
Its eigenvalue on any vector basis of the irreducible $J$-representation is 
$J(J+1)$.  \\
Moreover any state in the tensor product of two irreducible representations 
$R_{1} \otimes R_{2}$ is written in the crystal basis as one and only one 
tensor product of a $R_{1}$ state by a $R_{2}$ state \cite{Kashi}.  For 
example, taking for $R_{1}$ and $R_{2}$ the two-dimensional representation 
$J=\sfrac{1}{2}$ of ${\cal U}_{q \rightarrow 0}(sl(2))$ with states $\vert 
+ \rangle$ and $\vert - \rangle$, one will get in $R_{1} \otimes R_{2}$ the 
$J=1$ representation displayed by $\vert +,+ \rangle$, $\vert -,+ \rangle$ 
and $\vert -,- \rangle$, and the $J=0$ representation with the state $\vert 
+,- \rangle$.

Now let us state the main hypothesis of our model \cite{FSS}.

\subsection{The assumptions}
 
\noindent{\bf Assumption I} - The four nucleotides containing the bases: 
adenine $(A)$ and guanine $(G)$, deriving from purine, and cytosine $(C)$ 
and thymine $(T)$, coming from pyrimidine, are the basis vectors of a 
crystal basis of the $(1/2, 1/2)$ irreducible representation of the quantum 
algebra ${\cal U}_{q}(sl(2) \oplus sl(2))$ in the limit $q \rightarrow 0$.  
In the following, we denote with $\pm$ the basis vector corresponding to 
the eigenvalue $\pm 1/2$ of $J^{3}_{\alpha}$, where $\alpha = H \; (V)$ 
specifies the generator of the first (second) $sl(2)$.  We assume the 
following ``spin'' structure (we remind that the thymine $T$ in the DNA is 
replaced by the uracile $U$ in the RNA):
\begin{eqnarray} 
&sl(2)_{H}& \nonumber \\
C \equiv (+,+) &\qquad\longleftrightarrow\qquad& U \equiv (-,+) \nonumber 
\\
sl(2)_{V} \updownarrow && \updownarrow sl(2)_{V} \\
G \equiv (+,-) &\qquad\longleftrightarrow\qquad& A \equiv (-,-) \nonumber 
\\
&sl(2)_{H}& \nonumber
\end{eqnarray} 
Let us remark that the $H$-symmetry is associated to the purine-pyrimidine 
structure, while the $V$-symmetry reflects the complementarity rule (that 
is $A-T/U$ and $C-G$ interactions). \\

\medskip

\noindent {\bf Assumption II} - The \emph{codons} are the basis vectors, in 
the crystal basis, of the irreducible representations build up by the 
tensor product of three 4-dimensional $(\sfrac{1}{2},\sfrac{1}{2})$ 
fundamental representations describing the nucleotides.  \\
We have reported in Table \ref{tablerep} the assignment of the codons 
classified in the representations which appear in the r.h.s. of the 
following relation:
\begin{equation}
(\sfrac{1}{2},\sfrac{1}{2}) \;\otimes\; (\sfrac{1}{2},\sfrac{1}{2}) 
\;\otimes\; (\sfrac{1}{2},\sfrac{1}{2}) \; = \; (\sfrac{3}{2},\sfrac{3}{2}) 
\;\oplus\; 2\,(\sfrac{3}{2},\sfrac{1}{2}) \;\oplus\; 
2\,(\sfrac{1}{2},\sfrac{3}{2}) \;\oplus\; 4\,(\sfrac{1}{2},\sfrac{1}{2})
\end{equation}

In \cite{FSS}, an operator (called the \emph{reading or ribosome operator}) 
${\cal R}$ has been constructed out of the algebra ${\cal U}_{q \rightarrow 
0}(sl(2) \oplus sl(2))$, which describes the multiplet structure of the the 
genetic code in the following way: \emph{two codons have the same 
eigenvalue under ${\cal R}$ if and only if they are associated to the same 
amino-acid}.  Moreover an ``Hamiltonian'' depending on 4 parameters has 
been build up which gives a very satisfactory fit of the 16 values of the 
free energy released in the folding of a RNA sequence into a base paired 
double helix.

\medskip

Let us close this section by drawing the reader's attention to Fig. 
\ref{figA} where is specified for each codon its position in the 
appropriate representation.  The diagram of states for each representation 
is supposed to lie in a separate parallel plane.  Thick lines connect 
codons associated to the same amino-acid.  One remarks that each segment 
relates a couple of codons belonging to the same representation or to two 
different representations.  This last case occurs for quadruplets or 
sextets of codons associated to the same amino-acid.  It is the purpose of 
this letter to show a remarkable relation between such multiplets of codons 
(or amino-acids) involving the same subset of representation and (branching 
ratios of) the probabilities of presence of codons in the amino-acid 
biosynthesis.

\section {Correlation of codon usage}

We define the codon usage as the frequency of use of a given codon in the 
process of biosynthesis of all the amino-acids.  We define the probability 
of usage of the codon $XYZ$ of a given amino-acid as the ratio between the 
occurrence of the codon $XYZ$ and the occurrence $N$ of the corresponding 
amino-acid, i.e.  as the relative codon frequency, in the limit of very 
large $N$.  Here and in the following the labels $X,Y,Z,V$ represent the 
bases $C,U,G,A$.  The frequency rate of usage of a codon in a multiplet is 
connected to its probability of usage $P(XYZ \rightarrow \mbox{a.a.})$.  It 
is reasonable to assume that $P(XYZ \rightarrow \mbox{a.a.})$ depends on: 
-- the biological organism (b.o.)  from which the sequence considered has 
been extracted \\
-- the sequence analyzed \\
-- the nature of the neighboring codons in the sequence \\
-- the amino-acid (a.a.) \\
-- the nature and structure of the multiplet associated to the amino-acid 
\\
-- the biological environment \\
-- the properties of the codon itself ($XYZ$).

\medskip

We neglect the time in which the biosynthesis process takes place as we 
assume that the biosynthesis processes are considered at the same time, at 
least compared to the time scale of evolution of the genetic code.  We 
define the \emph{branching ratio} $B_{ZV}$ as
\begin{equation} 
B_{ZV} = \frac{P(XYZ \rightarrow \mbox{a.a.})}{P(XYV \rightarrow 
\mbox{a.a.})}
\label{eq8}
\end{equation}
We argue that in the limit of very large number of codons, for a fixed 
biological organism and amino-acid, the branching ratio depends essentially 
on the properties of the codon.  In our model this means that in this limit 
$B_{ZV}$ is a function, depending on the type of the multiplet, on the 
\emph{quantum numbers} of the codons $XYZ$ and $XYV$, i.e.  on the labels 
$J_{\alpha}, J^{3}_{\alpha}$ , where $\alpha = H$ or $V$, and on an other 
set of quantum labels leaving out the degeneracy on $J_{\alpha}$; in Table 
\ref{tablerep} different irreducible representations with the same values 
of $J_{\alpha}$ are distinguished by an upper label.  Moreover we assume 
that $B_{ZV}$, in the limit above specified, depends only on the 
irreducible representation (IR) of the codons, i.e.:
\begin{equation}   
B_{ZV} = F_{ZV}(b.o.; IR(XYZ); IR(XYV)) \label{eqbr}
\end{equation}  
Let us point out that the branching ratio has a meaning only if the codons 
$XYZ$ and $XYU$ are in the same multiplet, i.e.  if they code the same 
amino-acid.  \\
We consider the quartets and sextets.  There are five quartets and three 
sextets in the eukariotic code: that will allow a rather detailed analysis.  
Moreover the 3 sextets appear as the sum of a quartet and a doublet, see 
Table \ref{tablerep}.  In the following we consider only the quartet 
sub-part of the sextets.  We recall that the 5 amino-acids coded by the 
quartets are: [Pro, Ala, Thr, Gly ,Val] and the 3 amino-acids coded by the 
sextets are: [Leu, Arg, Ser].  There are, for the quartets, 6 branching 
ratios, of which only 3 are independent.  We choose as fundamental ones the 
ratios $B_{AG}$, $B_{CG}$ and $B_{UG}$.  It happens that we can define 
several functions $B_{ZV}$, considering ratios of probability of codons 
differing for the first two nucleotides $XY$, i.e.
\begin{eqnarray}
B_{ZV} & = & F_{ZV}(b.o.; IR(XYZ); IR(XYV)) \nonumber \\
B'_{ZV} & = & F_{ZV}(b.o.; IR(X'Y'Z); IR(X'Y'V))
\end{eqnarray}

Then if the codon $XYZ$ ($XYV$) and $X'Y'Z$ ($X'Y'V$) are respectively in 
the same irreducible representation, it follows that
\begin{equation}
B_{ZV} = B'_{ZV}
\end{equation}

The analysis was performed on a set of data retrieved from the data bank of 
``Codon usage tabulated from GenBank'' \cite{data}.  In particular we 
analyzed two different data set: the first one comprises all the data of at 
least 2 000 codons, while the second set represents all the data with at 
least 30 000 codons.  The referring organism for the analysis was 
\emph{Homo sapiens}, whose codon usage table derives from the analysis of 
more than 12 500 coding sequences, and corresponds to about 6 000 000 
codons.

\medskip

\begin{figure}
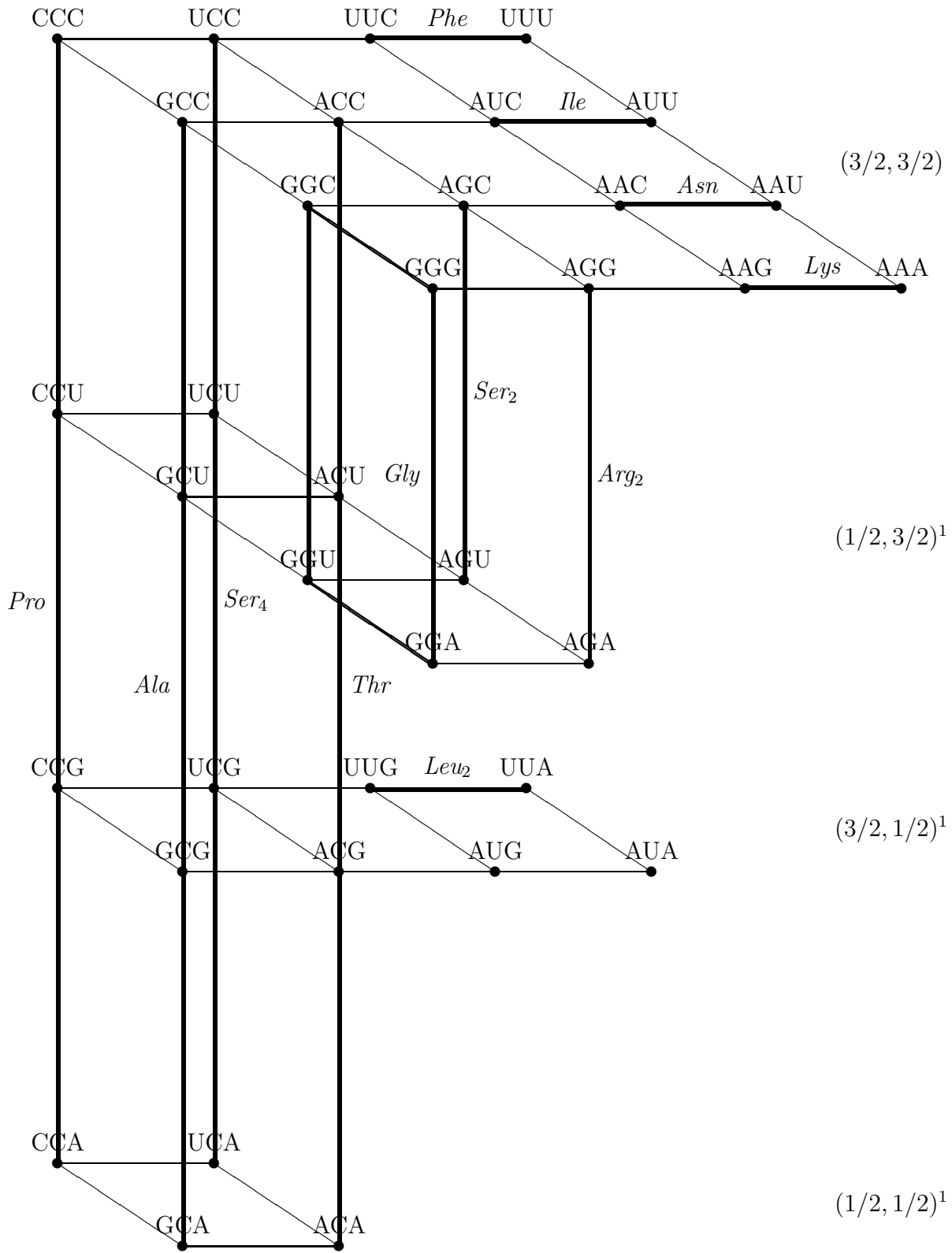

\caption{Classification of the codons in the different crystal bases.}
\label{figA}
\end{figure}

\begin{figure}
\caption{Branching ratio $B_{AG}$ for the vertebrate series.}
\label{fig1}
\includegraphics{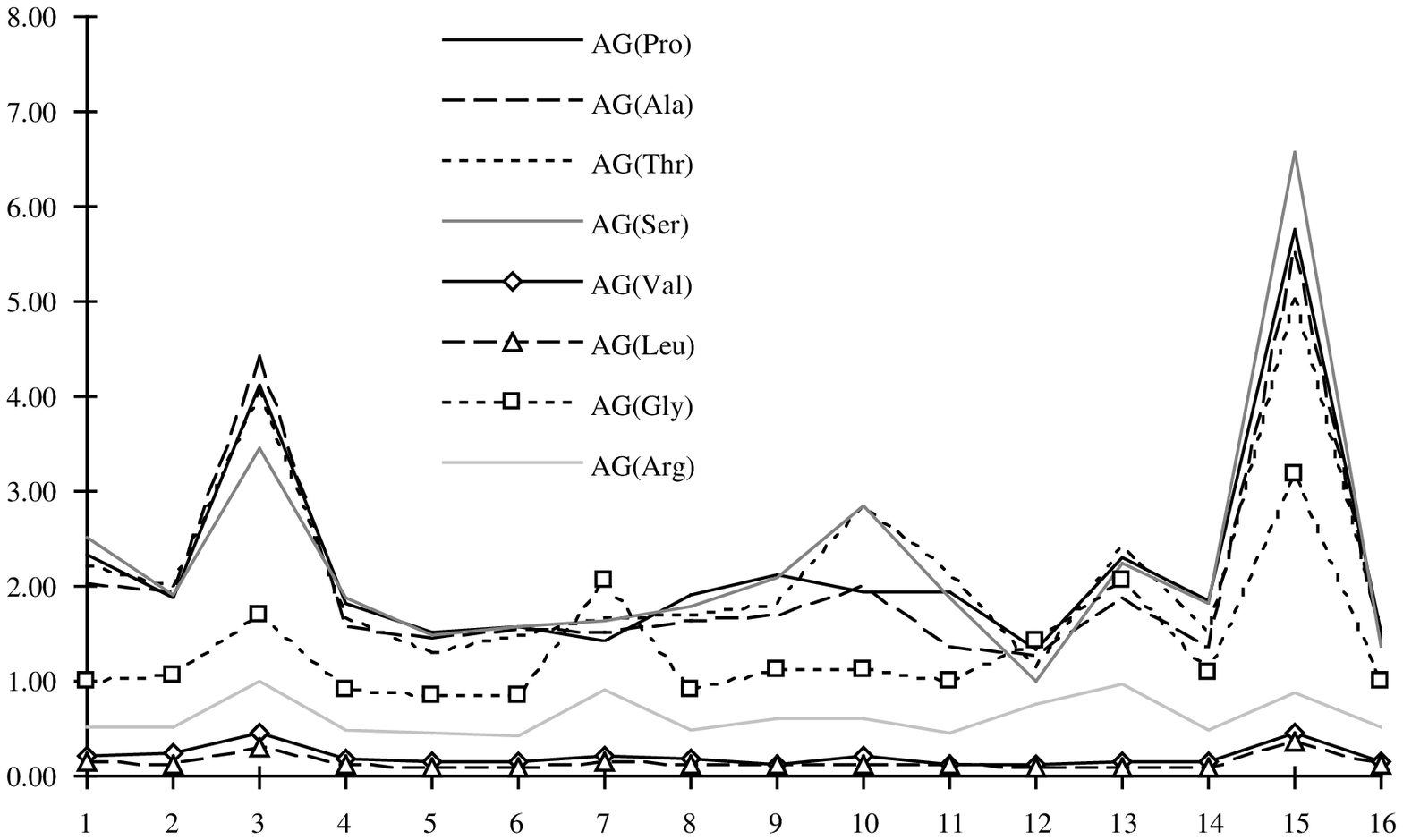}
\end{figure}

\begin{figure}
\caption{Branching ratio $B_{CG}$ for the vertebrate series.}
\label{fig2}
\includegraphics{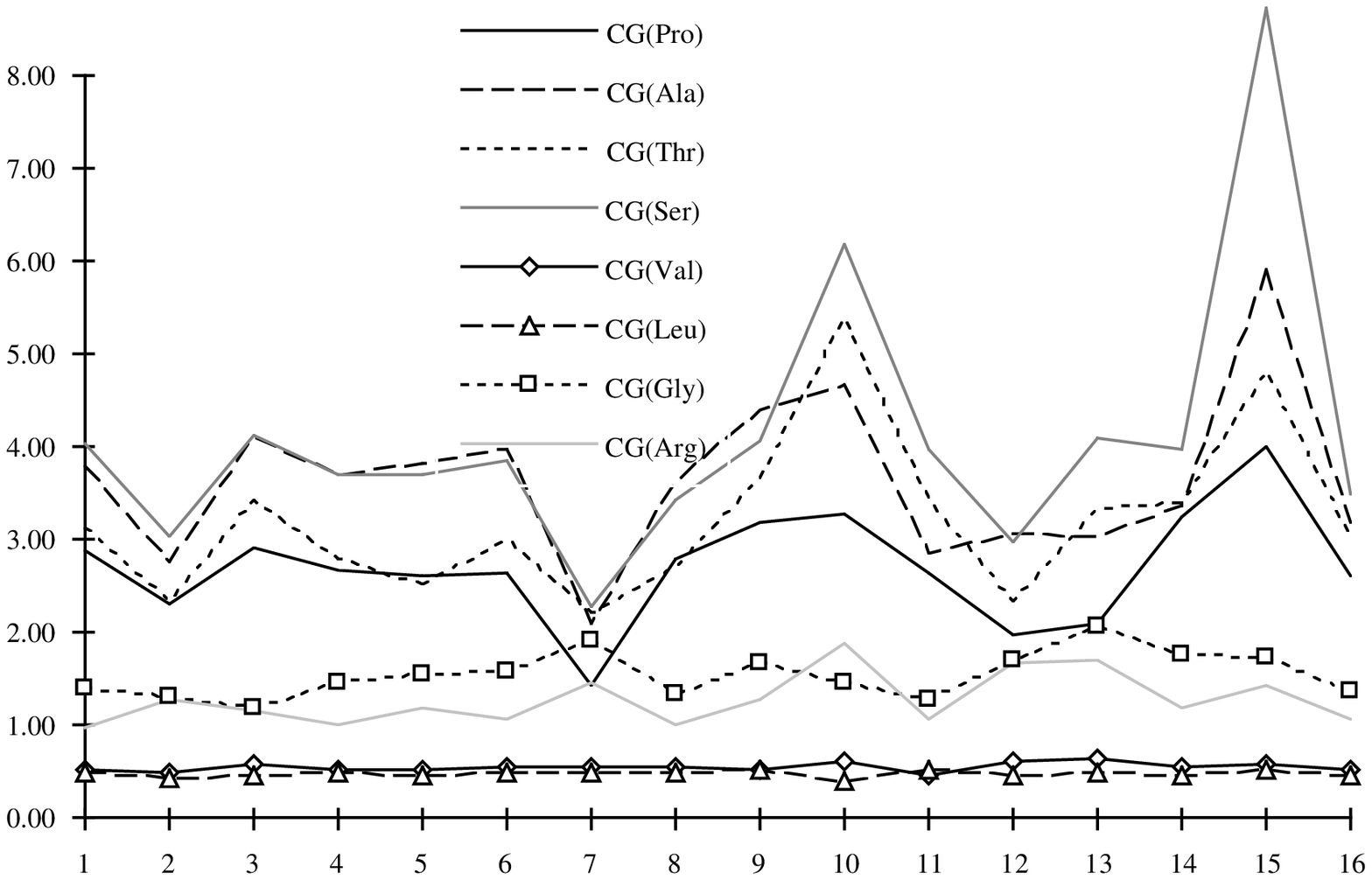}
\end{figure}

\begin{figure}
\caption{Branching ratio $B_{UG}$ for the vertebrate series.}
\label{fig3}
\includegraphics{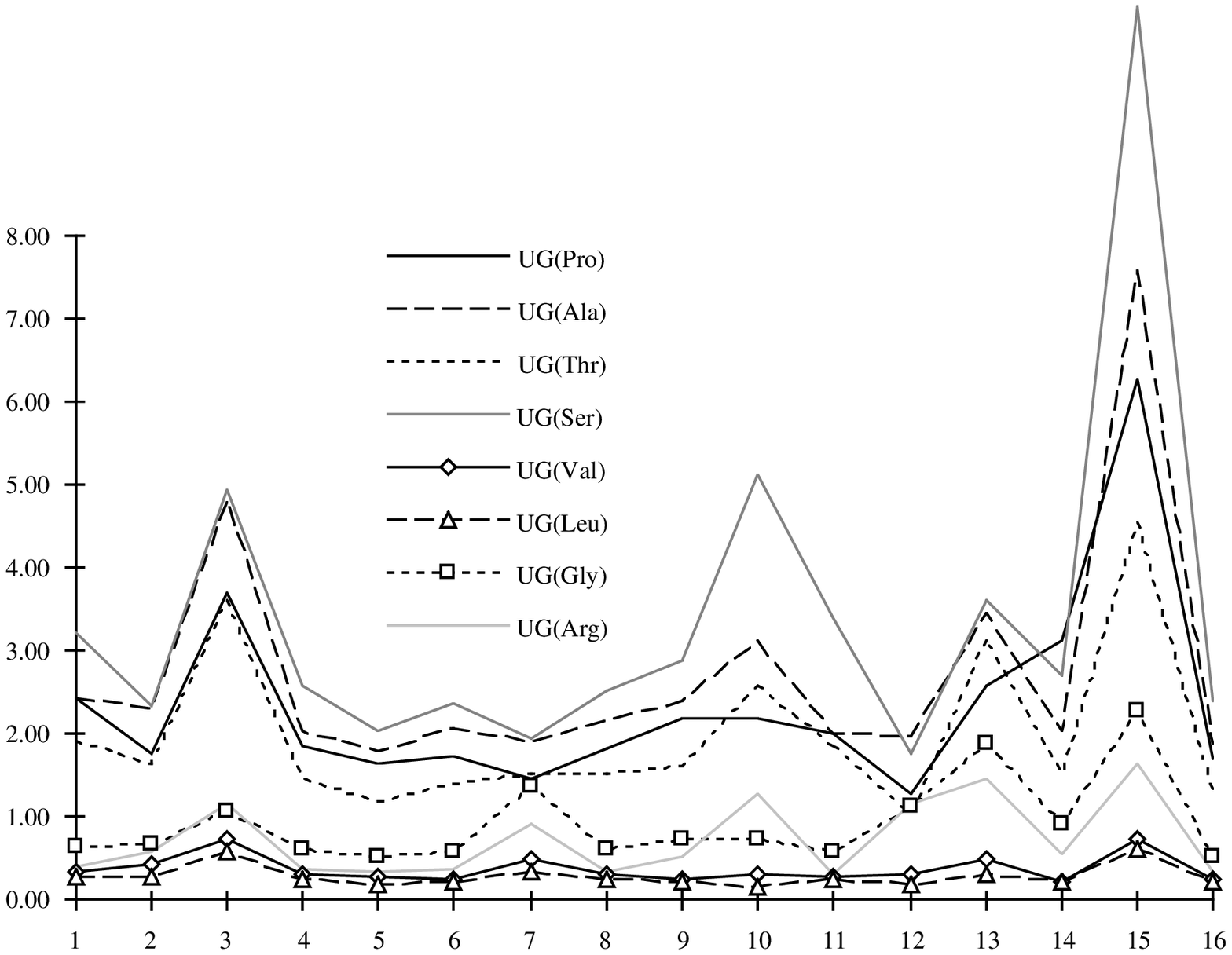}
\end{figure}

Three quartets, coding the amino-acids Pro, Ala and Thr, have exactly the 
same content in irreducible representations, see Table \ref{tablerep}.  In 
Table \ref{tabledata} we report the 16 biological organisms with highest 
statistics.  In Figs.  \ref{fig1}, \ref{fig2} and \ref{fig3} the $B_{AG}$ , 
$B_{UG}$ and $B_{CG}$ are reported for the 8 amino-acids coded by the 
quartets and sextets showing:
\begin{itemize}
\item 
a clear correlation between the four amino-acids Pro, Ala, Thr and Ser.  
{From} Table \ref{tablerep} we see that for these amino-acids the 
irreducible representation involved in the numerator of the branching 
ratios (see (\ref{eq8})) is always the same: $(1/2, 1/2)^1$ for $B_{AG}$, 
$(1/2, 3/2)^1$ for $B_{UG}$, $(3/2, 3/2)$ for $B_{CG}$, while the 
irreducible representation in the denominator is $(1/2, 1/2)^1$ for the 
whole set.  The relative position of each of these quartets of codons can 
be more easily visualized in Fig.  \ref{figA} where Pro, Ala, Thr and Ser 
(quartet part) constitute the four edges of a vertical column linking the 
representation $(1/2, 1/2)^1$, sitting at the ground floor, first to the 
representation $(3/2, 1/2)^1$, then to the $(1/2, 3/2)^1$ one and finally 
to the representation $(3/2, 3/2)$, this last one located at the top floor.
\item
a clear correlation between the two amino-acids Val and Leu.  {From} Table 
\ref{tablerep} we see that also for these two amino-acids the irreducible 
representation in the numerator of (\ref{eq8}) is the same: $(1/2, 1/2)^3$ 
for $B_{AG}$, $(1/2, 3/2)^2$ for $B_{UG}$, $(1/2, 3/2)^2$ for $B_{CG}$, and 
the irreducible representation in the denominator is $(1/2, 1/2)^3$.  
Considering Fig. \ref{figA}, it is now the two representations $(1/2, 
1/2)^3$ and $(1/2, 3/2)^2$ which are brought together, the codons 
associated to Val and Leu (quartet part) determining the vertices of two 
parallel and vertical plaquettes.
\item
no correlation of the Arg and also of the Gly with the others amino-acids, 
in agreement with the irreducible representation assignment of Table 
\ref{tablerep}.  Indeed we can note in Fig. \ref{figA} that the 
representations $(1/2, 1/2)^2$ and $(3/2, 1/2)^2$ are connected by the 
codon quartet relative to Arg and (but) only by this multiplet.  We also 
remark the Gly quartet in the representations $(1/2, 3/2)^1$ and $(3/2, 
3/2)$: its position is completely different from the above discussed 
quartets which show up in these representations.
\end{itemize}

Then in Figs.  \ref{fig4}, \ref{fig5} and \ref{fig6} we have drawn the 
normalized branching ratios $\widehat{B}_{PG}$, $P \in \{A,U,C\}$, defined 
by:
\begin{equation}
\widehat{B}_{PG} = \frac{B_{PG}}{\sum_{a.a.} \, B_{AG}} \label{eqnorm}
\end{equation}
where the sum $\sum_{a.a.}$ is extended to the eight amino-acids above 
listed.  The mean value and the standard deviation are:

\begin{center}
\begin{tabular}{|l|rrrrrrrr|}
\hline
& Pro & Ala & Thr & Ser & Val & Leu & Arg & Gly \\ \hline
$\langle\widehat{B}_{AG}\rangle$ & 1.60 & 1.46 & 1.57 & 1.61 & 0.16 & 0.11 
& 0.50 & 1.00 \\
$\sigma(\widehat{B}_{AG})$ & 0.16 & 0.16 & 0.17 & 0.21 & 0.03 & 0.02 & 0.15 
& 0.29 \\ \hline
$\langle\widehat{B}_{CG}\rangle$ & 1.24 & 1.66 & 1.46 & 1.83 & 0.26 & 0.23 
& 0.60 & 0.73 \\
$\sigma(\widehat{B}_{CG})$ & 0.15 & 0.18 & 0.15 & 0.23 & 0.05 & 0.04 & 0.16 
& 0.18 \\ \hline
$\langle\widehat{B}_{UG}\rangle$ & 1.49 & 1.71 & 1.24 & 2.07 & 0.25 & 0.19 
& 0.45 & 0.60 \\
$\sigma(\widehat{B}_{UG})$ & 0.26 & 0.13 & 0.14 & 0.32 & 0.06 & 0.04 & 0.22 
& 0.22 \\
\hline
\end{tabular}
\end{center}

These diagrams show an universal behaviour of $\widehat{B}_{PG}$ which has 
the same value independently of the biological organism.  We have omitted 
in the diagram the branching ratio of the amino-acid Gly as it is dependent 
from the branching ratios of the other amino-acids due to our definition 
eq.  (\ref{eqnorm}).  In our model this behaviour can easily be understood 
if the branching ratio $B_{ZV}$ has the factorized form
\begin{equation}
B_{ZV} = \Phi_{ZV}(b.o.)  \, \psi_{ZV}(IR(XYZ); IR(XYV)) \label{eqfa}
\end{equation} 
This factorization explains also the correlation in the behaviour between 
the values of $B_{PG}$ for different biological organisms, see Figs.  
\ref{fig1}, \ref{fig2} and \ref{fig3}.  Finally we report in the table 
below the mean value and the standard deviation for the case of biological 
organisms with low statistics to put in evidence the effects of the 
statistics.

\begin{center}
\begin{tabular}{|l|rrrrrrrr|}
\hline
& Pro & Ala & Thr & Ser & Val & Leu & Arg & Gly \\ \hline
$\langle\widehat{B}_{AG}\rangle$ & 1.77 & 1.49 & 1.67 & 1.13 & 0.17 & 0.15 
& 0.61 & 1.01 \\
$\sigma(\widehat{B}_{AG})$ & 0.67 & 0.40 & 0.49 & 0.56 & 0.09 & 0.18 & 0.34 
& 0.44 \\ \hline
$\langle\widehat{B}_{CG}\rangle$ & 1.29 & 1.55 & 1.52 & 1.82 & 0.25 & 0.23 
& 0.62 & 0.71 \\
$\sigma(\widehat{B}_{CG})$ & 0.47 & 0.39 & 0.41 & 0.53 & 0.08 & 0.08 & 0.32 
& 0.32 \\ \hline
$\langle\widehat{B}_{UG}\rangle$ & 1.50 & 1.61 & 1.26 & 2.09 & 0.25 & 0.19 
& 0.51 & 0.60 \\
$\sigma(\widehat{B}_{UG})$ & 0.58 & 0.39 & 0.39 & 0.64 & 0.10 & 0.09 & 0.28 
& 0.32 \\
\hline
\end{tabular}
\end{center}

\section{Conclusions}

The basic elements of our model of the genetic code are the 4 nucleotides 
and the 64 codons come out as composed states.  The symmetry algebra ${\cal 
U}_{q \rightarrow 0}(sl(2) \oplus sl(2))$ has two main characteristics.  
Firstly, it encodes the stereochemical property of a base confering quantum 
numbers to each nucleotide.  Secondly, it admits representation spaces with 
the remarkable property that the vector bases of the tensor product are 
ordered sequences of the basic elements (nucleotides).  The model does not 
necessarily assign the codons in a multiplet (in particular the quartets, 
sextets and triplet) to the same irreducible representation.  This feature 
is relevant.  Indeed, as we have shown in this paper, it may explain the 
correlation between the branching ratio of the codon usage of different 
codons coding the same amino-acid.  Let us remark that the assignments of 
the codons to the different irreducible representations is a 
straightforward consequence of the tensor product once assigned the 
nucleotides to the fundamental irreducible representation, see our first 
assumption.
 
\begin{figure}
\caption{Normalized branching ratio $B_{AG}$ for the vertebrate series.}
\label{fig4}
\includegraphics{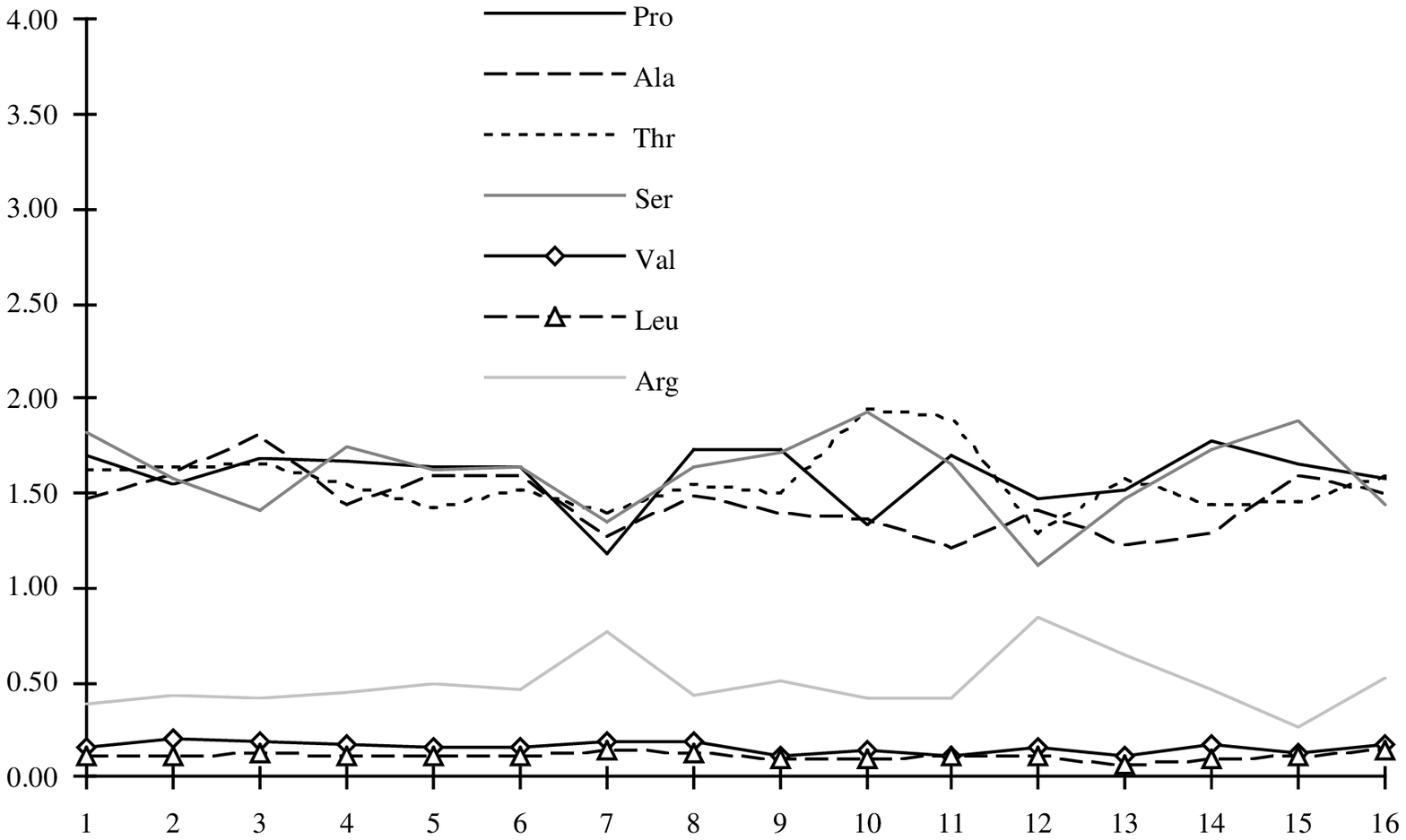}
\end{figure}

\begin{figure}
\caption{Normalized branching ratio $B_{CG}$ for the vertebrate series.}
\label{fig5}
\includegraphics{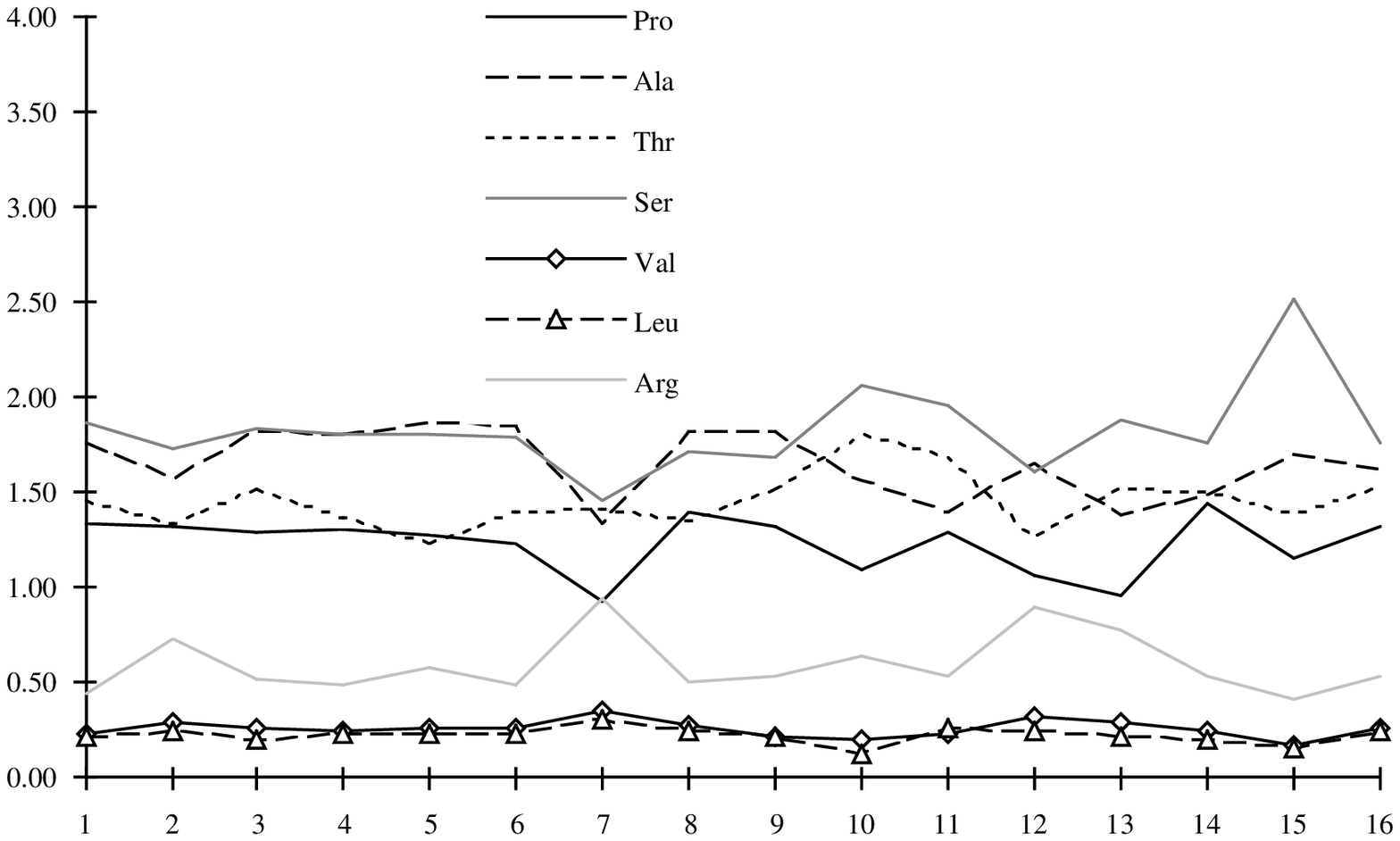}
\end{figure}

\begin{figure}
\caption{Normalized branching ratio $B_{UG}$ for the vertebrate series.}
\label{fig6}
\includegraphics{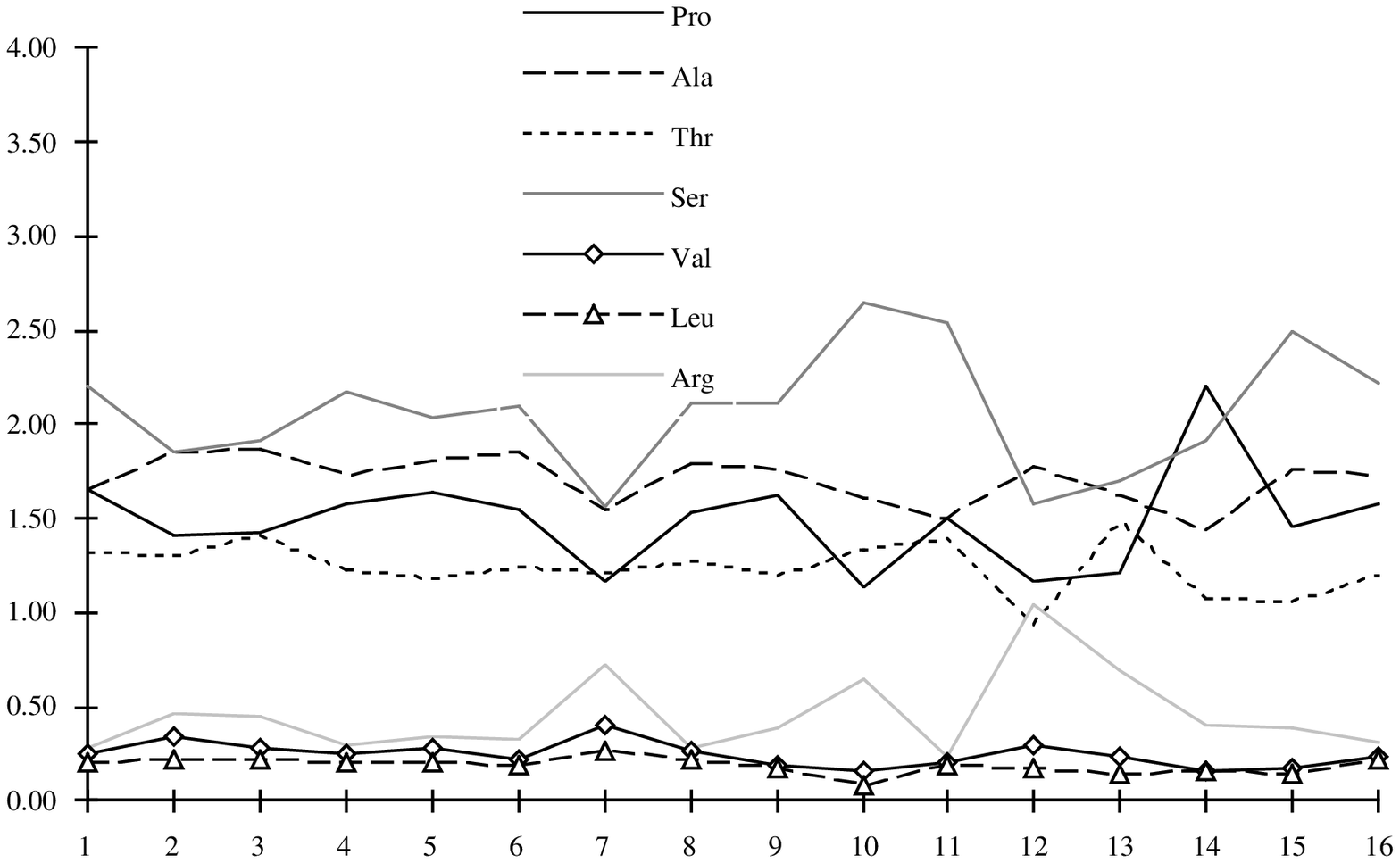}
\end{figure}

It is a prevision of our model that for \emph{any biological organism} 
belonging to the vertebrate series, in the limit of large number of 
biosynthetized amino-acids, the ratios $B_{AG}$, $B_{UG}$ and $B_{CG}$ for, 
respectively, Pro, Ala, Thr and Ser (Val and Leu) should be very close.  
Let us remark that obviously these ratios depend on the biological organism 
and we are unable to make any prevision on their values, but only that 
their values should be correlated.  Our analysis has also shown an 
universal behaviour of the normalized branching ratio of the codon usage 
for the vertebrates, which was not evidently expected in our model, but 
which can easily be explained assuming a factorized form for the $B_{ZV}$.  
So, assuming the factorization (\ref{eqfa}), we foresee that the normalized 
ratio $\widehat{B}_{AG}$, $\widehat{B}_{UG}$ and $\widehat{B}_{CG}$ should 
be given for any biological organism by the values reported in Figs.  
\ref{fig4}, \ref{fig5} and \ref{fig6}.

A first analysis including biological organisms belonging to the 
invertebrate and plant series show that the pattern of correlation is still 
present, even in a less striking way, but significant deviations appear for 
some biological organisms.  A more detailed analysis with extension to the 
other multiplets, in particular the doublets, and to other series of 
biological organisms will be done in a further more detailed publication.

\bigskip

\textbf{Acknowledgments} We are deeply indebted with Maria Luisa Chiusano 
for providing us the data which have allowed the analysis presented in this 
work and for very useful discussions.  It is also a pleasure to thank J.C. 
Le Guillou for discussions and encouragements.

\clearpage

% PICTURE 1
\begin{center}
\begin{picture}(400,-600)
% plateau 1
\multiput(0,0)(75,0){4}{\circle*{5}}
\multiput(0,0)(75,0){4}{\line(3,-2){60}}
\multiput(0,0)(75,0){3}{\line(1,0){75}}
\multiput(60,-40)(75,0){4}{\circle*{5}}
\multiput(60,-40)(75,0){4}{\line(3,-2){60}}
\multiput(60,-40)(75,0){3}{\line(1,0){75}}
\multiput(120,-80)(75,0){4}{\circle*{5}}
\multiput(120,-80)(75,0){4}{\line(3,-2){60}}
\multiput(120,-80)(75,0){3}{\line(1,0){75}}
\multiput(180,-120)(75,0){4}{\circle*{5}}
\multiput(180,-120)(75,0){3}{\line(1,0){75}}
\multiput(150,0)(60,-40){4}{\thicklines\line(1,0){75}}
\multiput(150,1)(60,-40){4}{\thicklines\line(1,0){75}}
\put(0,10){\makebox(0.4,0.6){CCC}}
\put(75,10){\makebox(0.4,0.6){UCC}}
\put(150,10){\makebox(0.4,0.6){UUC}}
\put(225,10){\makebox(0.4,0.6){UUU}}
\put(60,-30){\makebox(0.4,0.6){GCC}}
\put(135,-30){\makebox(0.4,0.6){ACC}}
\put(210,-30){\makebox(0.4,0.6){AUC}}
\put(285,-30){\makebox(0.4,0.6){AUU}}
\put(120,-70){\makebox(0.4,0.6){GGC}}
\put(195,-70){\makebox(0.4,0.6){AGC}}
\put(270,-70){\makebox(0.4,0.6){AAC}}
\put(345,-70){\makebox(0.4,0.6){AAU}}
\put(180,-110){\makebox(0.4,0.6){GGG}}
\put(255,-110){\makebox(0.4,0.6){AGG}}
\put(330,-110){\makebox(0.4,0.6){AAG}}
\put(405,-110){\makebox(0.4,0.6){AAA}}
% plateau 2
\multiput(0,-180)(60,-40){4}{\circle*{5}}
\multiput(0,-180)(60,-40){3}{\line(3,-2){60}}
\multiput(75,-180)(60,-40){4}{\circle*{5}}
\multiput(75,-180)(60,-40){3}{\line(3,-2){60}}
\multiput(0,-180)(60,-40){4}{\line(1,0){75}}
\put(0,-170){\makebox(0.4,0.6){CCU}}
\put(75,-170){\makebox(0.4,0.6){UCU}}
\put(60,-210){\makebox(0.4,0.6){GCU}}
\put(135,-210){\makebox(0.4,0.6){ACU}}
\put(120,-250){\makebox(0.4,0.6){GGU}}
\put(195,-250){\makebox(0.4,0.6){AGU}}
\put(180,-290){\makebox(0.4,0.6){GGA}}
\put(255,-290){\makebox(0.4,0.6){AGA}}
\put(120,-80){\thicklines\line(3,-2){60}}
\put(120,-81){\thicklines\line(3,-2){60}}
\put(120,-260){\thicklines\line(3,-2){60}}
\put(120,-261){\thicklines\line(3,-2){60}}
% plateau 3
\multiput(0,-360)(75,0){4}{\circle*{5}}
\multiput(0,-360)(75,0){4}{\line(3,-2){60}}
\multiput(0,-360)(75,0){3}{\line(1,0){75}}
\multiput(60,-400)(75,0){4}{\circle*{5}}
\multiput(60,-400)(75,0){3}{\line(1,0){75}}
\put(150,-360){\thicklines\line(1,0){75}}
\put(150,-361){\thicklines\line(1,0){75}}
\put(0,-350){\makebox(0.4,0.6){CCG}}
\put(75,-350){\makebox(0.4,0.6){UCG}}
\put(150,-350){\makebox(0.4,0.6){UUG}}
\put(225,-350){\makebox(0.4,0.6){UUA}}
\put(60,-390){\makebox(0.4,0.6){GCG}}
\put(135,-390){\makebox(0.4,0.6){ACG}}
\put(210,-390){\makebox(0.4,0.6){AUG}}
\put(285,-390){\makebox(0.4,0.6){AUA}}
% plateau 4
\multiput(0,-540)(75,0){2}{\circle*{5}}
\multiput(0,-540)(75,0){2}{\line(3,-2){60}}
\multiput(60,-580)(75,0){2}{\circle*{5}}
\multiput(0,-540)(60,-40){2}{\line(1,0){75}}
\put(0,-530){\makebox(0.4,0.6){CCA}}
\put(75,-530){\makebox(0.4,0.6){UCA}}
\put(60,-570){\makebox(0.4,0.6){GCA}}
\put(135,-570){\makebox(0.4,0.6){ACA}}
% lignes verticales
\multiput(0,0)(75,0){2}{\thicklines\line(0,-1){540}}
\multiput(1,0)(75,0){2}{\thicklines\line(0,-1){540}}
\multiput(60,-40)(75,0){2}{\thicklines\line(0,-1){540}}
\multiput(61,-40)(75,0){2}{\thicklines\line(0,-1){540}}
\multiput(120,-80)(75,0){2}{\thicklines\line(0,-1){180}}
\multiput(121,-80)(75,0){2}{\thicklines\line(0,-1){180}}
\multiput(180,-120)(75,0){2}{\thicklines\line(0,-1){180}}
\multiput(181,-120)(75,0){2}{\thicklines\line(0,-1){180}}
% acides-amines
\put(165,-210){\makebox(0.4,0.6){\textit{Gly}}}
\put(210,-170){\makebox(0.4,0.6){\textit{Ser$_{2}$}}}
\put(270,-210){\makebox(0.4,0.6){\textit{Arg$_{2}$}}}
\put(187,10){\makebox(0.4,0.6){\textit{Phe}}}
\put(247,-30){\makebox(0.4,0.6){\textit{Ile}}}
\put(307,-70){\makebox(0.4,0.6){\textit{Asn}}}
\put(367,-110){\makebox(0.4,0.6){\textit{Lys}}}
\put(-15,-270){\makebox(0.4,0.6){\textit{Pro}}}
\put(45,-310){\makebox(0.4,0.6){\textit{Ala}}}
\put(90,-270){\makebox(0.4,0.6){\textit{Ser$_{4}$}}}
\put(150,-310){\makebox(0.4,0.6){\textit{Thr}}}
\put(187,-350){\makebox(0.4,0.6){\textit{Leu$_{2}$}}}
% representations
\put(400,-60){\makebox(0.4,0.6){$(3/2,3/2)$}}
\put(400,-240){\makebox(0.4,0.6){$(1/2,3/2)^1$}}
\put(400,-380){\makebox(0.4,0.6){$(3/2,1/2)^1$}}
\put(400,-560){\makebox(0.4,0.6){$(1/2,1/2)^1$}}
\end{picture}
\end{center}

\clearpage

\centerline{Figure 1 (cont'd)}

% PICTURE 2
\begin{center}
\begin{picture}(400,-640)
% SUBPICTURE 1
% plateau 1
\multiput(0,0)(60,-40){4}{\circle*{5}}
\multiput(0,0)(60,-40){3}{\line(3,-2){60}}
\multiput(75,0)(60,-40){4}{\circle*{5}}
\multiput(75,0)(60,-40){3}{\line(3,-2){60}}
\multiput(0,0)(60,-40){4}{\thicklines\line(1,0){75}}
\multiput(0,1)(60,-40){4}{\thicklines\line(1,0){75}}
\put(0,10){\makebox(0.4,0.6){CUC}}
\put(75,10){\makebox(0.4,0.6){CUU}}
\put(60,-30){\makebox(0.4,0.6){GUC}}
\put(135,-30){\makebox(0.4,0.6){GUU}}
\put(120,-70){\makebox(0.4,0.6){GAC}}
\put(195,-70){\makebox(0.4,0.6){GAU}}
\put(180,-110){\makebox(0.4,0.6){GAG}}
\put(255,-110){\makebox(0.4,0.6){GAA}}
% plateau 2
\multiput(0,-180)(75,0){2}{\circle*{5}}
\multiput(0,-180)(75,0){2}{\line(3,-2){60}}
\multiput(60,-220)(75,0){2}{\circle*{5}}
\multiput(0,-180)(60,-40){2}{\thicklines\line(1,0){75}}
\multiput(0,-181)(60,-40){2}{\thicklines\line(1,0){75}}
\put(0,-170){\makebox(0.4,0.6){CUG}}
\put(75,-170){\makebox(0.4,0.6){CUA}}
\put(60,-210){\makebox(0.4,0.6){GUG}}
\put(135,-210){\makebox(0.4,0.6){GUA}}
% lignes verticales
\multiput(0,0)(75,0){2}{\thicklines\line(0,-1){180}}
\multiput(1,0)(75,0){2}{\thicklines\line(0,-1){180}}
\multiput(60,-40)(75,0){2}{\thicklines\line(0,-1){180}}
\multiput(61,-40)(75,0){2}{\thicklines\line(0,-1){180}}
% acides-amines
\put(-15,-90){\makebox(0.4,0.6){\textit{Leu$_{4}$}}}
\put(45,-130){\makebox(0.4,0.6){\textit{Val}}}
\put(157,-70){\makebox(0.4,0.6){\textit{Asp}}}
\put(217,-110){\makebox(0.4,0.6){\textit{Glu}}}
% SUBPICTURE 2
% plateau 1
\multiput(0,-300)(75,0){4}{\circle*{5}}
\multiput(0,-300)(75,0){4}{\line(3,-2){60}}
\multiput(0,-300)(75,0){3}{\line(1,0){75}}
\multiput(60,-340)(75,0){4}{\circle*{5}}
\multiput(60,-340)(75,0){3}{\line(1,0){75}}
\put(0,-290){\makebox(0.4,0.6){CGC}}
\put(75,-290){\makebox(0.4,0.6){UGC}}
\put(150,-290){\makebox(0.4,0.6){UAC}}
\put(225,-290){\makebox(0.4,0.6){UAU}}
\put(60,-330){\makebox(0.4,0.6){CGG}}
\put(135,-330){\makebox(0.4,0.6){UGG}}
\put(210,-330){\makebox(0.4,0.6){UAG}}
\put(285,-330){\makebox(0.4,0.6){UAA}}
\multiput(0,-300)(0,-180){2}{\thicklines\line(3,-2){60}}
\multiput(0,-301)(0,-180){2}{\thicklines\line(3,-2){60}}
\multiput(150,-300)(60,-40){2}{\thicklines\line(1,0){75}}
\multiput(150,-301)(60,-40){2}{\thicklines\line(1,0){75}}
% plateau 2
\multiput(0,-480)(75,0){2}{\circle*{5}}
\put(75,-480){\line(3,-2){60}}
\multiput(60,-520)(75,0){2}{\circle*{5}}
\multiput(0,-480)(60,-40){2}{\line(1,0){75}}
\put(0,-470){\makebox(0.4,0.6){CGU}}
\put(75,-470){\makebox(0.4,0.6){UGU}}
\put(60,-510){\makebox(0.4,0.6){CGA}}
\put(135,-510){\makebox(0.4,0.6){UGA}}
% lignes verticales
\multiput(0,-300)(75,0){2}{\thicklines\line(0,-1){180}}
\multiput(1,-300)(75,0){2}{\thicklines\line(0,-1){180}}
\put(60,-340){\thicklines\line(0,-1){180}}
\put(61,-340){\thicklines\line(0,-1){180}}
% acides-amines
\put(-15,-390){\makebox(0.4,0.6){\textit{Arg$_{4}$}}}
\put(90,-390){\makebox(0.4,0.6){\textit{Cys}}}
\put(187,-290){\makebox(0.4,0.6){\textit{Tyr}}}
\put(247,-330){\makebox(0.4,0.6){\textit{Ter}}}
\put(135,-350){\makebox(0.4,0.6){\textit{Trp}}}
% SUBPICTURE 3
% plateau
\multiput(0,-600)(75,0){2}{\circle*{5}}
\multiput(0,-600)(75,0){2}{\line(3,-2){60}}
\multiput(60,-640)(75,0){2}{\circle*{5}}
\multiput(0,-600)(60,-40){2}{\thicklines\line(1,0){75}}
\multiput(0,-601)(60,-40){2}{\thicklines\line(1,0){75}}
\put(0,-590){\makebox(0.4,0.6){CAC}}
\put(75,-590){\makebox(0.4,0.6){CAU}}
\put(60,-630){\makebox(0.4,0.6){CAG}}
\put(135,-630){\makebox(0.4,0.6){CAA}}
% acides-amines
\put(37,-590){\makebox(0.4,0.6){\textit{His}}}
\put(97,-630){\makebox(0.4,0.6){\textit{Gln}}}
% representations
\put(400,-60){\makebox(0.4,0.6){$(1/2,3/2)^2$}}
\put(400,-200){\makebox(0.4,0.6){$(1/2,1/2)^3$}}
\put(400,-320){\makebox(0.4,0.6){$(3/2,1/2)^2$}}
\put(400,-500){\makebox(0.4,0.6){$(1/2,1/2)^2$}}
\put(400,-620){\makebox(0.4,0.6){$(1/2,1/2)^4$}}
\end{picture}
\end{center}

\clearpage

\clearpage

\clearpage

\clearpage 

\begin{table}[t]
\caption{The eukariotic code. The upper label denotes different
IR.\label{tablerep}}
%\scriptsize
\footnotesize
\begin{center}
\begin{tabular}{|cc|cc|cc|cc|}
\hline
codon & a.a. & $J_{H}$ & $J_{V}$ & codon & a.a. & $J_{H}$ & $J_{V}$ \\
\hline
CCC & Pro & 3/2 & 3/2 & UCC & Ser & 3/2 & 3/2 \\
CCU & Pro & (1/2 & 3/2$)^1$ & UCU & Ser & (1/2 & 3/2$)^1$ \\
CCG & Pro & (3/2 & 1/2$)^1$ & UCG & Ser & (3/2 & 1/2$)^1$ \\
CCA & Pro & (1/2 & 1/2$)^1$ & UCA & Ser & (1/2 & 1/2$)^1$ \\
\hline
CUC & Leu & (1/2 & 3/2$)^2$ & UUC & Phe & 3/2 & 3/2 \\
CUU & Leu & (1/2 & 3/2$)^2$ & UUU & Phe & 3/2 & 3/2 \\
CUG & Leu & (1/2 & 1/2$)^3$ & UUG & Leu & (3/2 & 1/2$)^1$ \\
CUA & Leu & (1/2 & 1/2$)^3$ & UUA & Leu & (3/2 & 1/2$)^1$ \\ 
\hline
CGC & Arg & (3/2 & 1/2$)^2$ & UGC & Cys & (3/2 & 1/2$)^2$ \\
CGU & Arg & (1/2 & 1/2$)^2$ & UGU & Cys & (1/2 & 1/2$)^2$ \\
CGG & Arg & (3/2 & 1/2$)^2$ & UGG & Trp & (3/2 & 1/2$)^2$ \\
CGA & Arg & (1/2 & 1/2$)^2$ & UGA & Ter & (1/2 & 1/2$)^2$ \\
\hline
CAC & His & (1/2 & 1/2$)^4$ & UAC & Tyr & (3/2 & 1/2$)^2$ \\
CAU & His & (1/2 & 1/2$)^4$ & UAU & Tyr & (3/2 & 1/2$)^2$ \\
CAG & Gln & (1/2 & 1/2$)^4$ & UAG & Ter & (3/2 & 1/2$)^2$ \\
CAA & Gln & (1/2 & 1/2$)^4$ & UAA & Ter & (3/2 & 1/2$)^2$ \\
\hline
GCC & Ala & 3/2 & 3/2 & ACC & Thr & 3/2 & 3/2 \\
GCU & Ala & (1/2 & 3/2$)^1$ & ACU & Thr & (1/2 & 3/2$)^1$ \\
GCG & Ala & (3/2 & 1/2$)^1$ & ACG & Thr & (3/2 & 1/2$)^1$ \\
GCA & Ala & (1/2 & 1/2$)^1$ & ACA & Thr & (1/2 & 1/2$)^1$ \\
\hline
GUC & Val & (1/2 & 3/2$)^2$ & AUC & Ile & 3/2 & 3/2 \\
GUU & Val & (1/2 & 3/2$)^2$ & AUU & Ile & 3/2 & 3/2 \\
GUG & Val & (1/2 & 1/2$)^3$ & AUG & Met & (3/2 & 1/2$)^1$ \\
GUA & Val & (1/2 & 1/2$)^3$ & AUA & Ile & (3/2 & 1/2$)^1$ \\
\hline
GGC & Gly & 3/2 & 3/2 & AGC & Ser & 3/2 & 3/2 \\
GGU & Gly & (1/2 & 3/2$)^1$ & AGU & Ser & (1/2 & 3/2$)^1$ \\
GGG & Gly & 3/2 & 3/2 & AGG & Arg & 3/2 & 3/2 \\
GGA & Gly & (1/2 & 3/2$)^1$ & AGA & Arg & (1/2 & 3/2$)^1$ \\
\hline
GAC & Asp & (1/2 & 3/2$)^2$ & AAC & Asn & 3/2 & 3/2 \\
GAU & Asp & (1/2 & 3/2$)^2$ & AAU & Asn & 3/2 & 3/2 \\
GAG & Glu & (1/2 & 3/2$)^2$ & AAG & Lys & 3/2 & 3/2 \\
GAA & Glu & (1/2 & 3/2$)^2$ & AAA & Lys & 3/2 & 3/2 \\
\hline
\end{tabular}
\end{center}
\end{table} 

\vspace{-30mm}

\begin{table}[b]
\caption{Biological organisms with highest statistics.\label{tabledata}}
%\scriptsize
\footnotesize
\begin{center}
\begin{tabular}{|r|l|r|r|}
\hline
& Biological organism & number of sequences & number of codons \\
\hline
1 & Homo sapiens & 12 512 \phantom{--------} & 6 130 940 \phantom{-----} \\
2 & Gallus gallus & 1 319 \phantom{--------} & 638 532 \phantom{-----} \\
3 & Xenopus laevis & 1 144 \phantom{--------} & 493 437 \phantom{-----} \\
4 & Bos taurus & 1 182 \phantom{--------} & 478 270 \phantom{-----} \\
5 & Oryctolagus cuniculus & 639 \phantom{--------} & 321 129 \phantom{-----} \\
6 & Sus scrofa & 539 \phantom{--------} & 216 654 \phantom{-----} \\
7 & Danio rerio & 259 \phantom{--------} & 99 766 \phantom{-----} \\
8 & Canis familiaris & 230 \phantom{--------} & 94 444 \phantom{-----} \\
9 & Ovis aries & 275 \phantom{--------} & 81 177 \phantom{-----} \\
10 & Oncorhynchus mykiss & 128 \phantom{--------} & 42 794 \phantom{-----} \\
11 & Macaca mulatta & 110 \phantom{--------} & 34 510 \phantom{-----} \\
12 & Fugu rubripes & 63 \phantom{--------} & 32 943 \phantom{-----} \\
13 & Cyprinus carpio & 95 \phantom{--------} & 32 365 \phantom{-----} \\
14 & Equus caballus & 94 \phantom{--------} & 31 254 \phantom{-----} \\
15 & Rana cates beiana & 61 \phantom{--------} & 30 629 \phantom{-----} \\
16 & Felis catus & 83 \phantom{--------} & 30 031 \phantom{-----} \\
\hline
\end{tabular}
\end{center}
\end{table} 


\begin{thebibliography}{99}

\bibitem{FSS}
L. Frappat, A. Sciarrino, P. Sorba,
\textsl{A crystal basis for the genetic code,}
Preprint ENSLAPP-AL-671/97 and DSF-97/37, \texttt{physics/9801027}, to 
appear in Phys. Lett. A.

\bibitem{Kashi} 
M. Kashiwara, Commun. Math. Phys. \textbf{133} (1990) 249.

\bibitem{data}
Y. Nakamura, T. Gojobori, and T. Ikemura, Nucleic Acids Research 
\textbf{26} (1998) 334.

\end{thebibliography}
\end{document}